\documentclass[3p,times]{elsarticle}

\usepackage{ecrc}

\volume{00}

\firstpage{1}

\journalname{}

\runauth{Jun-Sik Sin, Song-Jin Im, Kwang-Il Kim}

\jid{}

\jnltitlelogo{}

\usepackage{amssymb}

\usepackage[figuresright]{rotating}

\begin{document}

\begin{frontmatter}



\dochead{}

\title{Asymmetric electrostatic properties of an electric double layer: a generalized Poisson-Boltzmann approach taking into account non-uniform size effects and water polarization}
\author{\large Jun-Sik Sin}
\large
\ead{js.sin@ryongnamsan.edu.kp}
\author{\large Song-Jin Im}
\author{\large Kwang-Il Kim}
\address{Department of Physics, \textbf{Kim Il Sung} University, Daesong District, Pyongyang, DPR Korea}

\begin{abstract}
\large

We theoretically study electrostatic properties of electric double layer using a generalized Poisson-Boltzmann approach taking into account the orientational ordering of water dipoles and the excluded volume effect of water molecules as well as those of positive and negative ions with different sizes in electrolyte solution.

Our approach enables one to predict that the number densities of water molecules, counterions and coions and the permittivity of electrolyte solution close to a charged surface, asymmetrically vary depending on both of sign and magnitude of the surface charge density and the volume of counterion.
We treat several phenomena in more detail.
Firstly, an increase in the volume of counterions and an increase in the surface charge density can cause the position of the minimum number density of water molecules to be farther from the charged surface. 
Secondly, width of the range of voltage in which the properties at the charged surface symmetrically vary decreases with increasing bulk salt concentration.
In addition, we show that the excluded volume effect of water molecules and the orientational ordering of water dipoles can lead to early onset and lowering of the maximum of electric capacitance according to surface voltage. Our approach and results can be applied to describing electrostatic properties of biological membranes and electric double layer capacitor for which excluded volume effects of water molecules and ions with different sizes may be important.
\end{abstract}

\end{frontmatter}


\section{Introduction}
\large
The concept of electric double layer was first presented by Herman von Helmholtz\cite%
{Helmholtz_annphys_1879} before more than one century. Since then many researchers have developed realistic theories of electric double layer to solve various problems of biology, medicine, colloid science and electrochemistry such as the binding of charged ligands to the membrane surface, the interactions of vesicles with the membrane, osteoblast attachment to biomaterials, fundamental nucleic acid processes, RNA folding, and differential capacitance of electric double layer capacitor \cite%
 {McLaughlin_annrev_1989, Smith_biomed_2004, Draper_annrev_2005, Wang_physchem_2009}.

To correctly represent electrostatic properties of electric double layer, a number of computational approaches such as Monte Carlo method and numerical solutions of integral equations\cite%
 {Boda_chemphys_2002,Lamperski_chemphys_2008,Jiang_chemphyslett_2011,Gonzalez_chemphys_1985}   were introduced, but they involve more complicated calculations than for the Poisson-Boltzmann (PB) approach.

However, the original PB approach proposed by Gouy and Chapmann \cite%
 {Gouy_physfrance_1910, Chapman_philos_1913}  doesn't consider the finite volumes of ions in electrolyte and it is known that the approach heavily overestimates the ionic concentrations close to charged surfaces in electrolyte. As an early attempt to eliminate such a shortcoming of the original PB approach, Stern \cite%
 {Stern_zelektrochem_1924} considered the finite size effect of ions by combining the Helmholtz model with the Gouy-Chapmann model. To include volume effects of ions directly into the PB approach, Bikerman\cite%
{Bikerman_philmag_1942} empirically modified Boltzmann distribution by correcting ion concentrations for the volume excluded by all ions.

In \cite%
 {Borukhov_prl_1997, Borukhov_electrochimica_2000,Iglic_physfran_1996, Bohnic_electrochimica_2001, Bohnic_bioelechem_2002, Bohnic_bioelechem_2005}, the authors considered finite volumes of ions and water molecules within lattice statistical mechanics approach. The assumption of the same size of ions and water molecules in electrolyte has been common to their work. Although this assumption does properly work in many situations, ionic transport across narrow channels and ionic adsorption in objects of subnanometer size cannot be described by means of the assumption. For example, when the size of a negative ion is larger than one of a positive ion, negative ionic transport across narrow channels of biological membranes or pores of electrode of electric double layer capacitor will hardly proceed.

Modified PB approaches using lattice statistics \cite%
{Chu_biophys_2007, Kornyshev_physchem_2007, Popovic_pre_2013}   were developed for considering the difference in sizes of positive and negative ions. For this purpose, in \cite%
 {Chu_biophys_2007}, a lattice in which one cell can contain several ions was used. In \cite%
 {Kornyshev_physchem_2007}, the authors semiempirically extended a modified PB equation to the case of an asymmetric salt. Recently, in \cite%
 {Popovic_pre_2013}, the authors assumed that the lattice cell size was an integer number of times smaller than a linear dimension of ion.

In fact, the Booth model \cite%
 {Booth_chemphys_1951,Booth_chemphys_1955} is well-known for generalization of the Onsager-Kirkwood-Fr\"ohlich permittivity model \cite%
 {Onsager_amchem_1936, Kirkwood_chemphys_1939} in the saturation regime of orientational ordering of water dipoles, but the model doesn't consider the sizes of both ions and water molecules in electrolyte solution.  

In \cite%
 {Iglic_bioelechem_2010, Gongadze_bioelechem_2012, Gongadze_bioelechem_2013,Gongadze_genphysiol_2013,Gongadze_electrochimica_2014}, the authors took into account the excluded volume effect of water molecules and the orientational ordering of water dipoles together with the excluded volume effect of ions in the modification of the PB approach using lattice statistics. In particular, the authors described that the permittivity of an electrolyte solution near a strongly charged surface may be heavily decreased by orientational ordering of water dipoles and depletion of water molecules  \cite%
 { Gongadze_bioelechem_2012, Gongadze_bioelechem_2013,Gongadze_genphysiol_2013}. In their approach, each particle occupies one cell of lattice based on the assumption that the ions and water molecules have the same excluded volume, which allowed the PB equations to have analytical and intuitive solutions. However, in cases where effects of difference in excluded volumes of water molecules and ions may be important, for example, where the electric capacitance is asymmetric due to the difference in sizes of positive and negative ions \cite%
 {Popovic_pre_2013}, a more general approach taking into account the effects is needed. 

In this paper we will incorporate not only the asymmetric size effect of ions in electrolyte but also both the orientational ordering of water dipoles and excluded volume effect of water molecules into the Poisson-Boltzmann approach. In a word, our approach generalizes that of \cite%
{Gongadze_bioelechem_2012} to include non-uniform ionic sizes. We introduce a lattice statistics where more than one cell can be occupied by each ion as in \cite%
 {Popovic_pre_2013} and also by each water molecule for considering effects of different excluded volumes of ions and water molecules. We show that electrostatic properties of electrolyte solution close to a charged surface aren't symmetric in positive and negative surface charge densities of the charged surface. We study effects of the volume of counterion and sign and magnitude of the surface voltage on electrostatic properties of the electrolyte solution. Finally, an early onset and lowering of the maximum of electric capacitance are predicted.

\section{The generalized Poisson-Boltzmann approach}
\large
We consider an electrolyte solution composed of multivalent ions and water molecules in contact with a charged planar surface, where a positive ion has charge $+ze_{0}$ and a negative ion has charge $-ze_{0}$.  The total free energy $F$ can be written in terms of the local electrostatic potential  $\psi\left(r\right)$ and the number densities of ions $c_{+}\left(r\right)$, $c_{-}\left(r\right)$ and water molecules
$c_{w}\left(r\right)=\left<\rho\left(\omega, r \right)\right>_{\omega}$.
 \begin{eqnarray}
	F=\int{d{\bf r}}\left(-\frac{\varepsilon_{0}\varepsilon E^2}{2}+e_{0}z\psi\left(c_{+}-c_{-}\right)+\left<\rho\left(\omega\right)\gamma{p_{0}}E\cos\omega\right>_{\omega}-\mu_{+}c_{+}-\mu_{-}c_{-}-\left<\mu_{w}\left(\omega\right)\rho\left(\omega\right)\right>_{\omega}-Ts\right),
\label{eq:7}
\end{eqnarray}
where $\left<f\left(\omega\right)\right>_{\omega}=\int f\left(\omega\right)2\pi\sin\left(\omega\right)  d\omega$ in which $\omega$ is the angle between the vector {\bf p} and the normal to the charged surface.  Here {\bf p} is the dipole moment of water molecules and {\bf E} is the electric field strength. The first term is the self energy of the electrostatic field, where $\varepsilon$ equals $n^2$ and $n=1.33$ is the refractive index of water. The next term corresponds to the electrostatic energy of the ions in the electrolyte solution, where $e_{0}$ is the elementary charge.
 The third one represents the electrostatic energy of water dipoles\cite%
{Gongadze_bioelechem_2012}, where  $\gamma=\left(2+n^2\right)/2$,  $p_{0}=\left|{\bf p}\right|$ and $E=\left|{\bf E}\right|$. The next three terms are responsible for coupling the system to a bulk reservoir, where $\mu_{+,-}$ are the chemical potentials of positive ions and negative ions and $\mu_{w}\left(\omega\right)$ is the chemical potential of water dipoles with orientational angle $\omega$.
$T$ is the temperature and $s$ is the entropy density.

Consider a unit volume of the electrolyte solution. The entropy density is the logarithm of the number of translational and orientational arrangements of non-interacting   $c_{+}$ positive ions,  $c_{-}$  negative ions and  $\rho\left(\omega_{i}\right)\Delta\Omega_{i}\left(i=1 \cdots N\right)$ water molecules, where $\Delta\Omega_{i}=2\pi  \sin\left(\omega_{i}\right) \Delta\omega$  is an element of a solid angle and $\Delta\omega=\pi/ N$. The positive ion, negative ion and water molecule occupy volumes of $V_{+},V_{-}$ and $V_{w}$, respectively.

Within a lattice statistics approach each particle in the solution occupies more than one cell of a lattice as in \cite%
 {Popovic_pre_2013}.
Considering translational arrangements of ions and orientational ordering of water dipoles, the number of arrangements can be calculated as follows. As in \cite%
{Popovic_pre_2013}, we first place $c_{+}$  positive ions of the volume $V_{+}$  and then  $c_{-}$  negative ones of the volume $V_{-}$  in the lattice. Finally, taking into account the orientational ordering of water dipoles, we put in  $\rho\left(\omega_{i}\right) \left(i=0,1,...\right)$ water molecules of the volume $V_{w}$   in the lattice. The number of arrangements $W$ is written as 
 \begin{equation}
	W=\frac{c_{s}\left(c_{s}-1\cdot v_{+}\right)\cdots\left(c_{s}-\left(c_{+}-1\right)v_{+}\right)}{c_{+}!}\frac{\left(c_{s}-c_{+}v_{\pm}\right)\cdots\left(c_{s}-c_{+}v_{\pm}-\left(c_{-}-1\right)v_{-}\right)}{c_{-}!}\frac{\left(c_{s}-c_{+}v_{+}-c_{-}v{-}\right)\cdots v_{w}}{\lim_{N\rightarrow\infty}\prod^{N}_{i=1}\rho\left(\omega_{i}\right)\Delta\Omega_{i}!},
\label{eq:1}
\end{equation}
where $v_{+,-,w}= V_{+,-,w}/a^3$ are the numbers of cells that the positive ion, negative ion and water molecule occupy, respectively. $c_{s}=1/a^3$ is the number of cells per unit volume and $a$ denotes the linear dimension of one cell.   

 From the standpoint of physics the entropy density should be symmetric in $+$ and $-$. For this purpose, we assume that the positive(negative) ion excludes $v_{\pm}=\left(v_{+}+v_{-}\right)/2$ for the negative(positive) ion.

 Expanding the logarithms of factorials using Stirling formula, we obtain the expression for the entropy density, $s=k_{B}\ln W$, 
 \begin{eqnarray}
	\frac{s}{k_{B}}=\ln W=-c_{+}\ln{a^3}-c_{-}\ln{a^3}-\left(\frac{1-c_{+}V_{+}-c_{-}V_{-}}{V_{w}}\right)\ln{a^3}-c_{+}\ln c_{+}-c_{-}\ln c_{-}-\left(\frac{1}{V_{+}}-c_{+}\right)\ln\left(1-c_{+}V_{+}\right)\nonumber\\+\left(\frac{1}{V_{-}}-\frac{c_{+}V_{\pm}}{V_{-}}\right)\ln\left(1-c_{+}V_{\pm}\right)-\left(\frac{1}{V_{-}}-\frac{c_{+}V_{\pm}}{V_{-}}-c_{-}\right)\ln\left(1-c_{+}V_{\pm}-c_{-}V_{-}\right)\nonumber\\-\left(\frac{1}{V_{w}}-\frac{c_{+}V_{+}}{V_{w}}-\frac{c_{-}V_{-}}{V_{w}}\right)\ln\left(1-c_{+}V_{+}-c_{-}V_{-}\right)-\lim_{N\rightarrow\infty}\sum^{N}_{i=1}\left[\rho\left(\omega_{i}\right)\Delta\Omega_{i} \ln\Delta\Omega_{i}+\rho\left(\omega_{i}\right)\Delta\Omega_{i} \ln\rho\left(\omega_{i}\right)-\rho\left(\omega_{i}\right)\Delta\Omega_{i}\right],
\label{eq:2}
\end{eqnarray}
where $k_{B}$  is the Boltzmann constant.
Assuming that $c_{i}V_{j}\left(i=+,-; j=+,-,\pm\right)$  are small, the above expression gets simpler:
\begin{eqnarray}
	\frac{s}{k_{B}}=-c_{+}\ln{c_{+}a^3}-c_{-}\ln{c_{-}a^3}-\left(\frac{1-c_{+}V_{+}-c_{-}V_{-}}{V_{w}}\right)\ln{a^3}+c_{+}\left(1-\frac{V_{+}}{V_{w}}\right)+c_{-}\left(1-\frac{V_{-}}{V_{w}}\right)\nonumber\\-\frac{c_{+}^{2}V_{+}}{2}-\frac{c_{-}^{2}V_{-}}{2}+\frac{\left(c_{+}V_{+}\right)^2}{2V_{w}}+\frac{\left(c_{-}V_{-}\right)^2}{2V_{w}}-c_{+}c_{-}V_{\pm}+c_{+}c_{-}\frac{V_{+}V_{-}}{V_{w}}-\lim_{N\rightarrow\infty}\sum^{N}_{i=1}\left[\rho\left(\omega_{i}\right)\ln\rho\left(\omega_{i}\right)+\rho\left(\omega_{i}\right)\ln\Delta\Omega_{i} \right]\Delta\Omega_{i}.
\label{eq:3}
\end{eqnarray}
Like the modification of the Poisson-Boltzmann approach of \cite%
{Popovic_pre_2013}, the expression for the entropy density is symmetric in negative and positive ions. 

 All lattice cells should be occupied by either ions or water molecules\cite%
{Iglic_physfran_1996, Gongadze_electrochimica_2014, Li_pre_2012, Li_pre_2011, Boschitsch_jcomchem_2012}, therefore
 \begin{eqnarray}
	c_{s}=c_{+}v_{+}+c_{-}v_{-}+c_{w}v_{w}.
\label{eq:0}
\end{eqnarray}

In order to find the free energy in equilibrium taking into account Eq.   
(\ref{eq:0}), we will use the method of undetermined multipliers. The Lagrangian of the electrolyte solution is 
 \begin{eqnarray}
	L=F-\int\lambda\left({\bf r}\right)\left(1-c_{+}V_{+}-c_{-}V_{-}-c_{w}V_{w}\right)d{\bf r},
\label{eq:8}
\end{eqnarray}
where $\lambda$ is a local Lagrange parameter.
Once the Lagrangian is established, the Euler$-$Lagrange equations are obtained and solved with respect to the functions $c_{+}, c_{-}$ and $\rho\left(\omega\right)$.
The variation of the Lagrangian with respect to $c_{+}$  yields an equation from which we can get the number density of positive ions in the electrolyte solution:
 \begin{eqnarray}
	\frac{\delta L}{\delta c_{+}}=e_{0}z\psi-\mu_{+}+k_{B}T\left(\ln c_{+}a^3+c_{+}V_{+}\left(1-\frac{V_{+}}{V_{w}}\right)+c_{-}V_{\pm}-c_{-}\frac{V_{+}V_{-}}{V_{w}}\right)+\lambda V_{+}=0.
\label{eq:9}
\end{eqnarray}
Using the boundary conditions $\psi\left(x\rightarrow\infty\right)=0$ and $c_{+,-}\left(x\rightarrow\infty\right)=c_{0}$ and $\lambda\left(x\rightarrow\infty\right)=\lambda_{0}$, we get the chemical potential for positive ions from Eq.(\ref{eq:9}):
 \begin{eqnarray}
	\mu_{+}=k_{B}T\left(\ln c_{0}a^3+c_{0}V_{+}\left(1-\frac{V_{+}}{V_{w}}\right)+c_{0}V_{\pm}-c_{0}\frac{V_{+}V_{-}}{V_{w}}\right)+\lambda_{0} V_{+}.
\label{eq:10}
\end{eqnarray}
Inserting Eq.(\ref{eq:10}) into Eq.(\ref{eq:9}), we obtain $c_{+}$   by exponentiation:
 \begin{eqnarray}
	c_{+}=c_{0}\exp\left(-\frac{e_{0}z\psi}{k_{B}T}\right)\exp\left(-\left[\left(c_{+}-c_{0}\right)V_{+}\left(1-\frac{V_{+}}{V_{w}}\right)+\left(c_{-}-c_{0}\right)\left(V_{\pm}-\frac{V_{+}V_{-}}{V_{w}}\right)\right]\right)\exp\left(\lambda V_{+}\right),
\label{eq:11}
\end{eqnarray}
where $\left(\lambda-\lambda_{0}\right)/k_{B}T\rightarrow\lambda$ for simplicity.
Like the derivation of Eq.(\ref{eq:11}), the expressions for  $c_{-}$ and $\rho\left(\omega\right)$ are simply obtained:
 \begin{eqnarray}
	c_{-}=c_{0}\exp\left(\frac{e_{0}z\psi}{k_{B}T}\right)\exp\left(-\left[\left(c_{-}-c_{0}\right)V_{-}\left(1-\frac{V_{-}}{V_{w}}\right)+\left(c_{+}-c_{0}\right)\left(V_{\pm}-\frac{V_{+}V_{-}}{V_{w}}\right)\right]\right)\exp\left(\lambda V_{-}\right),
\label{eq:12}
\end{eqnarray}
 \begin{eqnarray}
	\rho\left(\omega\right)=\rho_{0}\exp\left(-\frac{\gamma p_{0}E\cos\omega}{k_{B}T}\right)\exp\left(\lambda V_{w}\right).
\label{eq:13}
\end{eqnarray}
 In general, within our approach the number densities of ions and water molecules are obtained implicitly not explicitly.

In the case when the ions and water molecules have the same size and the surface charge density is negative, that is,  when $V_{+}=V_{-}=V_{w}$ and $\sigma<0$, we can recover all basic equations of \cite%
{Gongadze_bioelechem_2012}.  When we neglect orientational ordering of water dipoles, our approach is identical to that of \cite%
{Popovic_pre_2013}.

The Euler$-$Lagrange equation for  $\psi\left(r\right)$  yields the Poisson- Boltzmann equation
\begin{eqnarray}
	\nabla\left(\varepsilon_{0}\varepsilon_{r}\nabla\psi\right)=-e_{0}z\left(c_{+}-c_{-}\right),	
\label{eq:14}
\end{eqnarray}
where
 \begin{eqnarray}
	\varepsilon_{r} \equiv n^2+\frac{{\bf P}}{\varepsilon_{0}{\bf E}}.
\label{eq:4}
\end{eqnarray}
 Here, ${\bf P}$ is the polarization vector due to a total orientation of point-like water dipoles. 
From the planar symmetry of this problem, one can see that the electric field strength is perpendicular to the charged surface and have the same magnitude at all points equidistant from the charged surface. The $x$ axis points in the direction of the bulk solution and is perpendicular to the charged surface. Consequently, along the $x$ axis {\bf E} and {\bf P}  have only an $x$ component and {\bf P}  is given as \cite%
{Gongadze_bioelechem_2012}
\begin{eqnarray}
	{\bf P}\left(x\right)=c_{w}\left(x\right)\left(\frac{2+n^2}{3}\right)p_{0}\mathcal{L}\left(\gamma{p_{0}}E\beta \right)\hat{{\bf e}}, 
\label{eq:6}
\end{eqnarray}
where a function $\mathcal{L}\left(u\right)=\coth\left(u\right)-1/u$ is the Langevin function, $\hat{{\bf e}}={{\bf E}/E}$ and $\beta=1/\left(k_{B}T\right)$ .

Differentiation of Eqs.(\ref{eq:11})-(\ref{eq:13}) with respect to the distance from the charged surface provides linear algebraic equations in terms of $d{c_{+}}/dx,d{c_{-}}/dx,d{c_{w}}/dx,d{\lambda}/dx$:
 \begin{eqnarray}
	\frac{d{c_{+}}}{dx}=c_{+}\left[-\frac{d\Phi}{dx}-V_{+}\left(1-\frac{V_{+}}{V_{w}}\right)\frac{d{c_{+}}}{dx}-\left(V_{\pm}-\frac{V_{+}V_{-}}{V_{w}}\right)\frac{d{c_{-}}}{dx}+V_{+}\frac{d\lambda}{dx}\right], 
\label{eq:15}
\end{eqnarray}
 \begin{eqnarray}
	\frac{d{c_{-}}}{dx}=c_{-}\left[\frac{d\Phi}{dx}-V_{-}\left(1-\frac{V_{-}}{V_{w}}\right)\frac{d{c_{-}}}{dx}-\left(V_{\pm}-\frac{V_{+}V_{-}}{V_{w}}\right)\frac{d{c_{+}}}{dx}+V_{-}\frac{d\lambda}{dx}\right], 
\label{eq:16}
\end{eqnarray}
 \begin{eqnarray}
	\frac{d{c_{w}}}{dx}=c_{w}\left[\mathcal{L}\left(\gamma{p_{0}}E\beta\right)\left(\gamma{p_{0}}\beta\right)\frac{dE}{dx}+V_{w}\frac{d\lambda}{dx}\right], 
\label{eq:17}
\end{eqnarray}
where $\Phi\left(x\right)=e_{0}z\psi\left(x\right)/k_{B}T$.

Solving the system of Eqs.(\ref{eq:0}),(\ref{eq:15})-(\ref{eq:17}) for $d{c_{+}}/dx,d{c_{-}}/dx,d{c_{w}}/dx,d{\lambda}/dx$  results in the following coupled differential equations:
 \begin{eqnarray}
	\frac{d{c_{+}}}{dx}=\frac{c_{+}}{D}\left(c_{-}V_{-}^2+c_{-}V_{+}V_{-}-c_{-}c_{w}V_{+}V_{-}V_{w}-c_{-}c_{w}V_{-}^2 V_{w}+c_{-}c_{w}V_{-}V_{w}^2+c_{-}c_{w}V_{\pm}V_{w}^2+c_{w}V_{w}^2\right)\frac{d\Phi}{dx}\nonumber\\+\frac{c_{+}c_{w}V_{w}}{D}\left(-c_{-}V_{\pm}V_{-}+V_{+}+c_{-}V_{+}V_{-}\right){\mathcal{L}}\left(\gamma{p_{0}}E\beta\right)\left(\gamma{p_{0}}\beta\right)\frac{dE}{dx}, 
\label{eq:19}
\end{eqnarray}
 \begin{eqnarray}
	\frac{d{c_{-}}}{dx}=\frac{c_{-}}{D}\left(-c_{+}V_{+}^2-c_{+}V_{+}V_{-}+c_{+}c_{w}V_{+}V_{-}V_{w}+c_{+}c_{w}V_{+}^2 V_{w}-c_{+}c_{w}V_{+}V_{w}^2-c_{+}c_{w}V_{\pm}V_{w}^2-c_{w}V_{w}^2\right)\frac{d\Phi}{dx}\nonumber\\+\frac{c_{-}c_{w}V_{w}}{D}\left(-c_{+}V_{\pm}V_{+}+V_{-}+c_{+}V_{+}V_{-}\right)
\mathcal{L}\left(\gamma{p_{0}}E\beta\right)\left(\gamma{p_{0}}\beta\right)\frac{dE}{dx}, 
\label{eq:20}
\end{eqnarray}
 \begin{eqnarray}
	\frac{d{c_{w}}}{dx}=-\frac{c_{w}V_{w}}{D}\left[c_{+}V_{+}-c_{-}V_{-}+c_{+}c_{-}V_{\pm}\left(V_{+}-V_{-}\right)\right]\frac{d\Phi}{dx}\nonumber\\-\frac{c_{w}}{D}\left[c_{+}V_{+}^2+c_{-}V_{-}^2+c_{+}c_{-}V_{+}V_{-}\left(V_{+}+V_{-}-2V_{\pm}\right)\right]
\mathcal{L}\left(\gamma{p_{0}}E\beta\right)\left(\gamma{p_{0}}\beta\right)\frac{dE}{dx},
\label{eq:21}
\end{eqnarray}

where
 \begin{eqnarray}
	D=-c_{+}V_{+}^2\left(1-c_{w}V_{w}\right)\left(1+c_{-}V_{-}\right)-c_{-}V_{-}^2\left(1-c_{w}V_{w}\right)\left(1+c_{+}V_{+}\right)                                                             \nonumber\\-c_{w}V_{w}^2\left[\left(1+c_{+}V_{+}\right)\left(1+c_{-}V_{-}\right)-c_{+}c_{-}V_{\pm}^{2}\right]+2c_{+}c_{-}\left(1-c_{w}V_{w}\right)V_{+}V_{-}V_{\pm},
\label{eq:22}
\end{eqnarray}

The electrostatic potential and the number densities of the ions and water molecules are obtained by solving Eqs.(\ref{eq:14}),(\ref{eq:19})-(\ref{eq:22}).
In many cases of biological and chemical systems, effective volumes of positive ions, negative ions and water molecules in electrolyte solutions differ from each other, so that our approach will be useful for actual cases.

\section{Results and Discussion}
\large
Under the boundary conditions $\psi\left(x\rightarrow\infty\right)=0$  and $E\left(x=0\right)=\sigma/\left(\varepsilon_{0}\varepsilon_{r}\left(x=0\right)\right)$, we combine Eq.(\ref{eq:14}) and Eqs.(\ref{eq:19}-\ref{eq:22}) and solve these differential equations for $c_{+}, c_{-}, c_{w}, \psi$  by using the fourth order Runge-Kutta method. As in \cite%
{Gongadze_bioelechem_2012}, the water dipole moment $p_{0}$ should be 3.1 D(Debye is $3.336\times10^{-30}$C/m) so that far away from the charged surface the relative permittivity of electrolyte solution reaches 78\cite%
{Iglic_bioelechem_2010, Gongadze_bioelechem_2012}.
In calculations, we choose  $c_{0w}/N_{A}=55mol/l$ for the number density of water molecules in the bulk electrolyte solution \cite%
{Iglic_bioelechem_2010, Gongadze_bioelechem_2012}.

\begin{figure}
\includegraphics[width=1\textwidth]{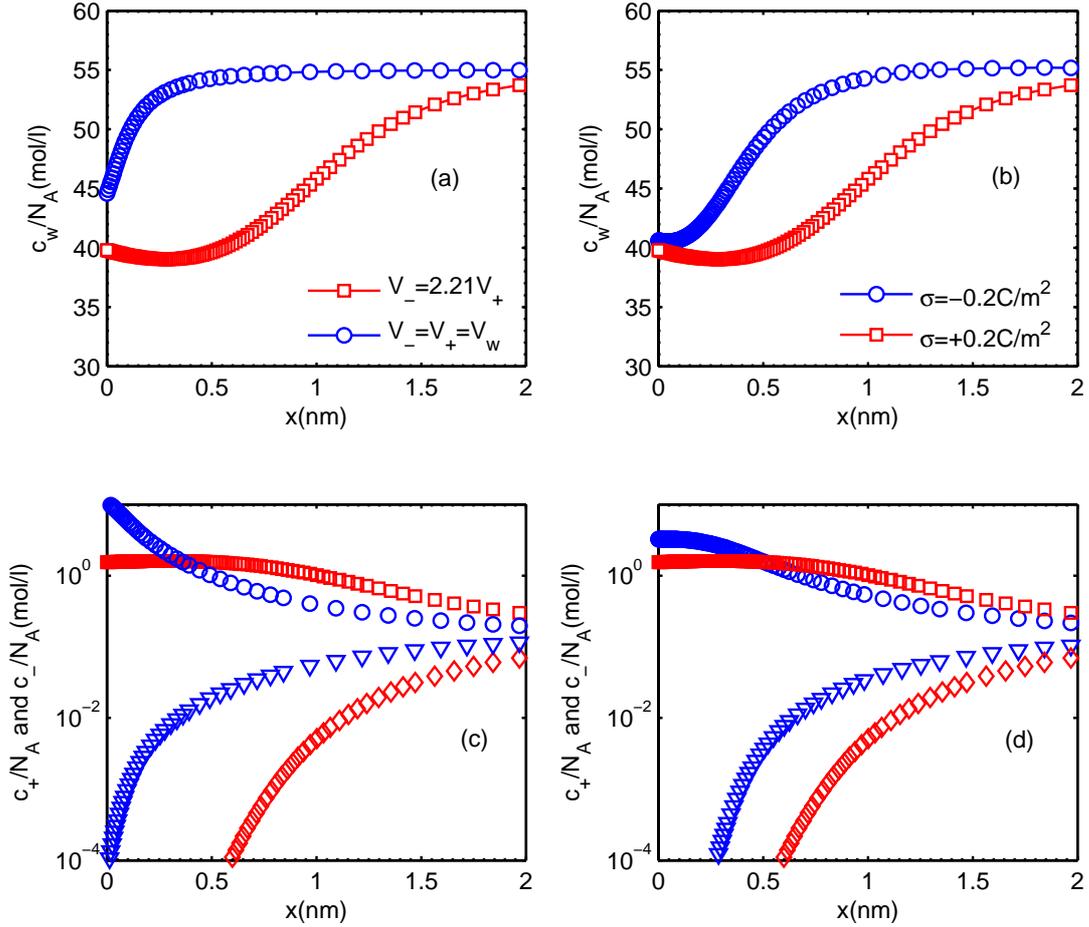}
\caption{(Color online) The number densities of water molecules, coions and counterions as functions of the distance from the charged planar surface.
Dipole moment of water $p_{0} = 3.1D$, monovalent bulk salt concentration $c_{0}/N_{A}= 0.15mol/l$ and temperature $T = 300K$.
(a) The number density of water molecules as a function of the distance from the charged planar surface. The case of $V_{-}=V_{+}=V_{w}=0.03nm^3$ is compared with the case of $V_{-}=2.21V_{+},V_{+}=0.15nm^3$. Here the surface charge density is $\sigma=+0.2C/m^2$.
(b). The number density of water molecules as a function of the distance from the charged planar surface. For $V_{-}=2.21V_{+},V_{+}=0.15nm^3$, the case of $\sigma=+0.2C/m^2$ is compared with the case of $\sigma=-0.2C/m^2$.
(c). For the case of $V_{-}=2.21V_{+},V_{+}=0.15nm^3$, the number densities of counterions (squares) and coions(diamonds).
      For the case of $V_{-}=V_{+}=V_{w}=0.03nm^3$, the number densities of counterions (circles) and coions(triangles).
      Here the surface charge density is $\sigma=+0.2C/m^2$.
(d). For the case of $\sigma=+0.2C/m^2$, the number densities of counterions (squares) and coions(diamonds).
      For the case of $\sigma= -0.2C/m^2$, the number densities of counterions (circles) and coions(triangles).
      Here  $V_{-}=2.21V_{+},V_{+}=0.15nm^3$.}
\label{fig:1}
\end{figure}

\begin{figure}
\includegraphics[width=1\textwidth]{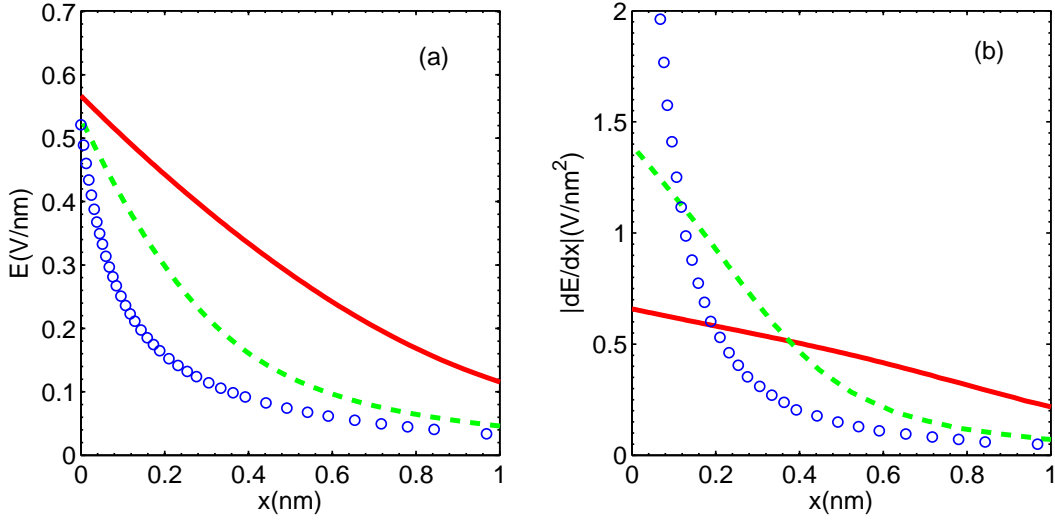}
\caption{(Color online) The magnitude of electric field strength (a)  and the magnitude of the first derivative of electric field strength (b) according to the distance from the charged surface for the cases of $V_{-}=V_{+}=V_{w}=0.03nm^3$(circles), $V_{-}=2.21V_{+},V_{+}=0.15nm^3 $ (solid line)and $V_{+}=2.21V_{-},V_{-}=0.15nm^3$(dashed line). Here surface charge density is $\sigma=+0.2C/m^2$ and other parameters are the same as in Fig. \ref{fig:1}(a).}
\label{fig:2}
\end{figure}

Fig. \ref{fig:1}(a) compares the calculated number densities of water molecules for $V_{-}=V_{+}=V_{w}=0.03nm^3$ used in \cite%
{Gongadze_genphysiol_2013} and $V_{+}=0.15nm^3, V_{-}=2.21V_{+}, V_{w}=0.03nm^3 $  (of our case), where the surface charge density is $\sigma=+0.2C/m^2$. We note that in \cite%
{Popovic_pre_2013} the asymmetric parameters $V_{+}=0.15nm^3, V_{-}=2.21V_{+}$ are used, however the authors didn't consider the excluded volume effect of water molecules.  Here,  monovalent bulk salt concentration is taken to be $c_{0}/N_A=0.15mol/\l$. In the case when ions and water molecules are of equal size the number density of water molecules is monotonously decreased towards the charged surface in agreement with \cite%
{Gongadze_genphysiol_2013}. For our case when ions and water molecules aren't of the same size, the number density of water molecules first decreases with increasing distance from the charged surface and reaches a minimum. Then this density increases and reaches the bulk value of water 55mol/l.  Such a behaviour of the number density of water molecules in the vicinity of the charged surface was also predicted and shown to be a consequence of a balance between the counterion Boltzman factor $c_{0}\exp\left(e_{0}|\phi\left(x\right)|\beta\right)$ and rotationally averaged water Boltzmann factor $c_{0w}\left<\exp\left(-\gamma p_{0}E\beta\cos\omega\right)\right>_{\omega}$ in \cite%
{Gongadze_genphysiol_2013}. However, it is noticed that in our case the surface charge density for this characteristic behaviour is smaller than that of \cite%
{Gongadze_genphysiol_2013}. This is attributed to the excluded volume of counterions larger than that in \cite%
{Gongadze_genphysiol_2013}. Fig. \ref{fig:1}(b) represents the number densities of water molecules for surface charge densities of equal magnitude and opposite sign, where $V_{+}=0.15nm^3, V_{-}=2.21V_{+}, V_{w}=0.03nm^3 $. It is shown that the position of the minimum for $\sigma=+0.2c/m^2$ is farther from the charged surface than that for $\sigma=-0.2c/m^2$. Beyond the positions of the minima, the number density of water molecules for $\sigma=+0.2c/m^2$ is smaller than that for $\sigma=-0.2c/m^2$. We note that the effective volume of counterion for $\sigma=+0.2c/m^2$ is larger than that for $\sigma=-0.2c/m^2$, which results in surface-charge-sign-dependent $c_{w}$ as mentioned above. 

 Fig. \ref{fig:1}(c) and Fig. \ref{fig:1}(d)  show the number densities of ions for the cases of Fig. \ref{fig:1}(a) and Fig. \ref{fig:1}(b), respectively.  Fig. \ref{fig:1}(c) and Fig. \ref{fig:1}(d) indicate that the number density of coions is much less than the number density of counterions near the charged surface.

As shown in Fig. \ref{fig:1}(b), it is interesting that an increase in effective volume of counterion causes the position of minimum number density of water molecules to be farther from the charged surface.  Eq.(\ref{eq:21}) provides a way to understand physical meaning of the fact. The value of the right term of Eq.(\ref{eq:21}) is equal to zero at the position of the minimum number density of water molecules.
Considering that the number density of coions near the charged surface is much smaller than that of counterions as shown in Fig. \ref{fig:1}(c) and Fig. \ref{fig:1}(d), Eq.(\ref{eq:23}) is obtained.
 \begin{eqnarray}
	\frac{V_{w}}{V_{-}}=\frac{\gamma{p_{0}}}{e_{0}z}\frac{\mathcal{L}\left(\gamma{p_{0}}E\beta)\right)}{E}\left|\frac{dE}{dx}\right|,
\label{eq:23}
\end{eqnarray}
where without loss of generality, we assume that the charged surface has a positive charge.
 But it is difficult to analyze this equation directly because all the physical quantities(i.e. the electric field strength, the number densities of water molecules and ions and the permittivity of the electrolyte solution) are subtly linked together. The electric field strength and the first derivative of the electric field strength are the key to understanding the above equation.
Fig. \ref{fig:2}(a) and Fig. \ref{fig:2}(b) show the magnitudes of the electric field strength and first derivative of electric field strength according to the distance from the charged surface, respectively, where the surface charge density is $\sigma=+0.2C/m^2$.  Other parameters are the same as in Fig. \ref{fig:1}(a).
Fig. \ref{fig:2}(a) and Fig. \ref{fig:2}(b) obviously illustrate the fact that the larger the volume of counterion, the weaker the screening property of counterion: an increase in volume of counterion causes  both an increase in the electric field strength and a decrease in the first derivative of electric field strength. We can easily know that the function $\mathcal{L}\left(x\right)/x$ in Eq.(\ref{eq:4}) decreases with increasing value of $x$. 
From Eq. (\ref{eq:23}) and the above facts, it is deduced that the position shift of minimum water density is attributed to difference in screening properties of counterions due to the difference in volumes of counterions.  

\begin{figure}
\includegraphics[width=0.5\textwidth]{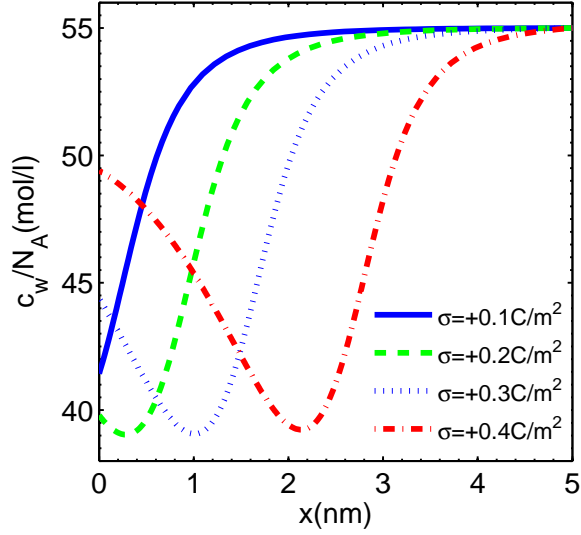}
\caption{(Color online)The number density of water molecules according to the distance from the charged surface  for different surface charge
densities: $\sigma = +0.1C/m^2,+0.2C/m^2,+0.3C/m^2,+0.4C/m^2$. Here $V_{-}=2.21V_{+},V_{+}=0.15nm^3$. Other parameters are the same as in Fig. \ref{fig:1}(a).}
\label{fig:6}
\end{figure}

Fig. (\ref{fig:6}) shows the variation of number density of water molecules with the distance from the charged surface for different surface charge densities.  Fig. (\ref{fig:6}) indicates that the minimum number density of water molecules doesn't depend on the surface charge density and that the larger the surface charge density, the more distant the position of the minimum value from the charged surface. In fact, an increase in surface charge density of the charged surface causes increases in magnitude for both the electric field strength and the first derivative of electric field strength in the electrolyte solution. Consequently, the position at which Eq.(\ref{eq:23}) is satisfied is farther from the charged surface than that for the original surface charge density.
\begin{figure}
\includegraphics[width=1\textwidth]{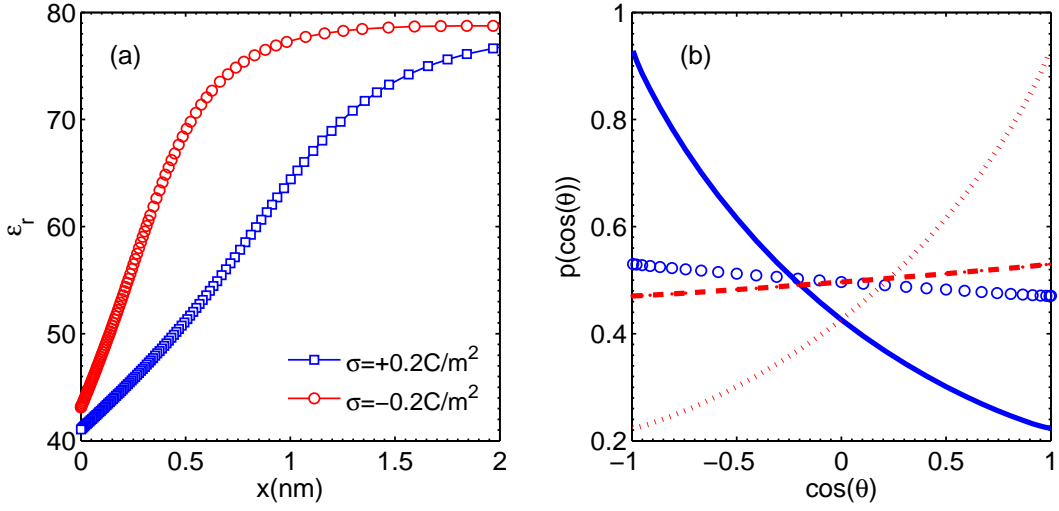}
\caption{(Color online) (a) The variation of the permittivity of the electrolyte solution with the distance from the charged surface for both surface charge
densities $\sigma = +0.2C/m^2$ and $\sigma = -0.2C/m^2$.
(b) The orientational distribution function of cos($\theta$) where $\theta$ is the angle between a water dipole and the vector normal to the electrode surface (pointing into the bulk). For the case of $\sigma = +0.2C/m^2$  the curves at the charged surface($x=0$)(dotted line) and $x=1nm$(dashed line)  are compared with corresponding ones at the charged surface($x=0$)(solid line) and $x=1nm$(circles) for the case of $\sigma =-0.2C/m^2$.
Other parameters are the same as in Fig. \ref{fig:1}(b). }
\label{fig:3}
\end{figure}

Fig. \ref{fig:3}(a) represents the permittivities of the electrolyte solution for $\sigma=+0.2c/m^2$ and $\sigma=-0.2c/m^2$. Other parameters are the same as in Fig. \ref{fig:1}(b). Fig. \ref{fig:3}(a) shows that the permittivity for the larger size of counterion is smaller than that for the smaller size of counterion. 
The reason for it is explained by two effects.
  On one hand,  the number density of water molecules for the larger size of counterion is small compared to that for the smaller size of counterion as shown in Fig. \ref{fig:1}(b). On the other hand, the increase in electric field strength due to difference in volumes of counterion(as shown in Fig. \ref{fig:2}(a)) also causes a decrease of permittivity of electrolyte solution since the function $\mathcal{L}\left(x\right)/x$ in Eq.(\ref{eq:4}) decreases with increasing value of $x$.  By using Eqs.(\ref{eq:4}), (\ref{eq:6}) together with the above facts, it is found that the permittivity of the electrolyte solution for the larger size of counterion becomes smaller than that for the smaller size of counterion. 
\begin{figure}
\includegraphics[width=1\textwidth]{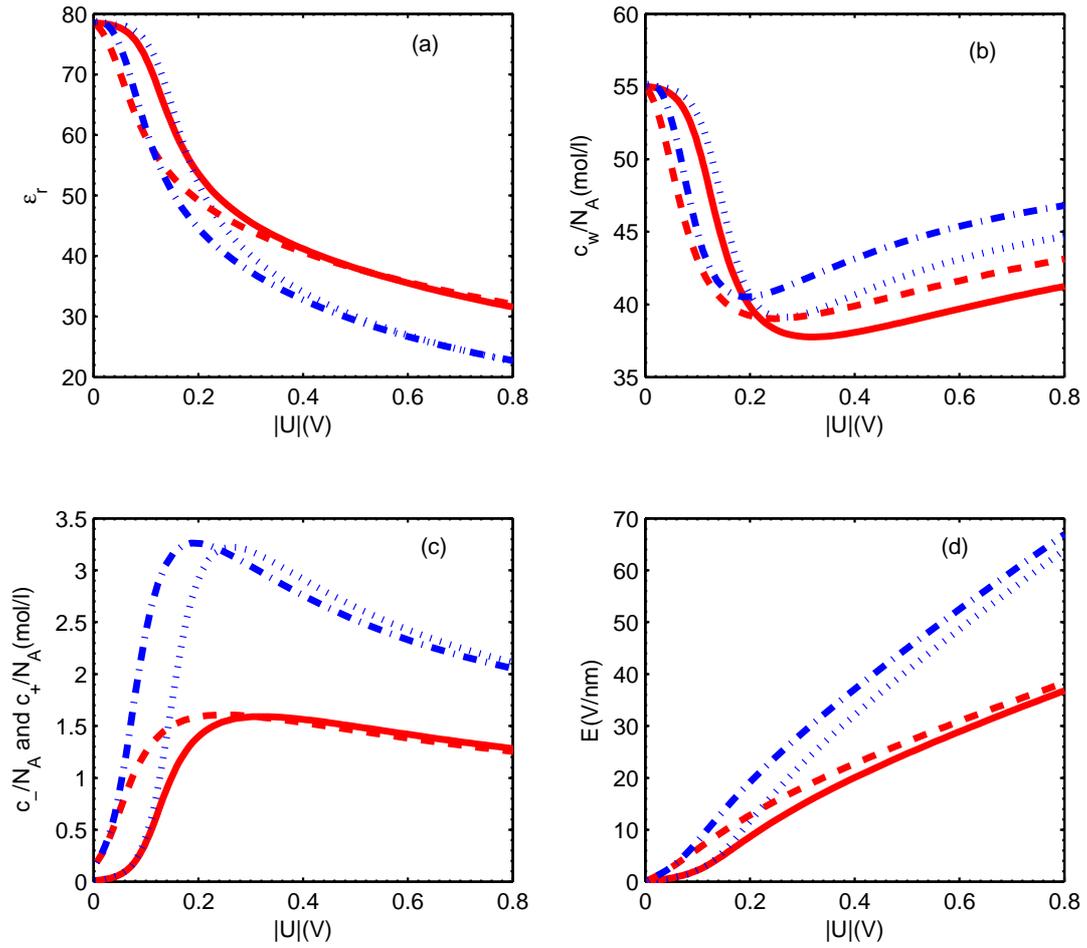}
\caption{(Color online)  Permittivity (a), number density of water molecules (b),  number densities of coions and counterions (c) and magnitude of electric field strength (d) in the electrolyte solution at charged planar surface according to magnitude of the surface voltage. 
Physical quantity for $c_{0}/N_{A}=0.15mol/l$ and positive voltage (dashed line), Physical quantity for $c_{0}/N_{A}=0.15mol/l$ and negative voltage(dash-dotted line), Physical quantity for $c_{0}/N_{A}=0.01mol/l$ and positive voltage(solid line), Physical quantity for $c_{0}/N_{A}=0.01mol/l$ and negative voltage(dotted line). Here $V_{-}=2.21V_{+},V_{+}=0.15nm^3$. Other parameters are the same as in Fig.  \ref{fig:1}(a). }
\label{fig:4}
\end{figure}
In order to understand the dependence of the permittivity on the distance from the charged surface, it is helpful to look at orientational distribution function $p\left(\Omega\right)$. According to the definition of orientational distribution function, $p\left(\Omega\right)d{\Omega}$ is the probability of finding a molecule with solid angles $\Omega=\left(\theta,\phi\right)$. In the case of planar surface,  $p\left(\Omega\right)$ is independent of the azimuthal angle $\phi$. Since $p\left(\Omega\right)$ should be normalized, $\int^{\pi}_{0}p\left(\cos\left(\theta\right)\right)\sin\left(\theta\right){d}\theta=1$. The orientational distribution function is obtained by applying Eq.(\ref{eq:13}) to the normalization condition. 
\begin{eqnarray}
	p\left(\cos\left(\theta\right)\right)=\frac{\gamma p_{0}E}{ k_{B}T}\cdot\frac{\exp\left(-\frac{\gamma p_{0}E\cos\left(\theta\right)}{k_{B}T}\right)}{\left(\exp\left(\frac{\gamma p_{0}E}{k_{B}T}\right)-\exp\left(-\frac{\gamma p_{0}E}{k_{B}T}\right)\right)}
\label{eq:24}
\end{eqnarray}
Fig. \ref{fig:3}(b) shows the orientational distribution functions  $p\left(\cos\left(\theta\right)\right)$ at different distances($x=0$ and $x=1nm$) from the charged surface for $\sigma=+0.2c/m^2$ and $\sigma=-0.2c/m^2$. Other parameters are the same as in Fig. \ref{fig:3}(a).
 Due to simplicity of our mean-field approach, Fig.\ref{fig:3}(b) differs from the results of \cite%
{willard_faraday_2009, spohr_electrochim_1999, siepmann_jchemphys_1995}. However, Fig. \ref{fig:3}(b) well represents the fact that at the positively charged surface the population of water dipoles pointing into the charged surface($\cos\left(\theta\right)<0$) is depleted but at the negatively charged surface a large population of water dipoles points into the charged surface. 
Since a large population of water molecules is highly oriented under a large electric field, more polarization is hardly provided and the permittivity of electrolyte solution should therefore decrease.  In \cite%
{quiroga_jelechemsoc_2014}, the authors reported both the electric field strength dependence and spatial variation of relative permittivity for different solvent composition. The behaviours of permittivities for our and their approaches are similar. Unlike the situation in our approach, the authors assumed that possible directions for the solvent dipoles are perpendicular to the electrode surface: the dipoles moments pointing toward the surface or opposite to the surface.

Fig. \ref{fig:4}(a), Fig. \ref{fig:4}(b), Fig. \ref{fig:4}(c) and Fig. \ref{fig:4}(d) show the permittivity,  the number density of water molecules, the number densities of coions and counterions, and the magnitude of the electric field strength at the charged surface as functions of surface voltage for different bulk salt concentrations ($c_{0}/N_{A}=0.01mol/l,0.15mol/l$), respectively. Other parameters are the same as in Fig. \ref{fig:1}(a). 

Let's consider the case of  $c_{0}/N_{A}=0.01mol/l$.
For $|U|<0.1V$, the symmetric shape of the permittivity curve of the electrolyte is due to the small number density of counterions near the charged surface and the negligible excluded volume effect of counterions as shown in Fig. \ref{fig:4}(c).  For $|U|>0.1V$, the permittivity doesn't symmetrically vary with $U$ due to the different excluded volume effects of positive and negative ions. In the region of $0.1<|U|<0.2V$, the permittivity of the electrolyte solutions for the larger size of counterion is larger than that for the smaller size of counterion. Conversely, for $|U|>0.2V$, the permittivity for the smaller size of counterion is higher than for the larger size of counterion. 
The behaviour of the permittivity for $|U|>0.2V$ is attributed to the following two facts. 
Firstly, Fig. \ref{fig:4}(b) shows that at the charged surface, the number density of water molecules is a non-monotonic function according to the surface charge density. This fact is understood by the formation of the minimum number density of water molecules as shown in Fig. \ref{fig:1}(a). Secondly, Fig. \ref{fig:4}(d) shows that an increase in magnitude of the surface voltage at the charged surface is accompanied by an increase in magnitude of the electric field strength at the position.  Finally, the magnitudes of both quantities for the larger size of counterion are smaller than that for the smaller size of counterion as shown in Fig. \ref{fig:4}(b) and Fig. \ref{fig:4}(d). The behaviour of the permittivity for $|U|>0.2V$ is explained by using  Eqs.(\ref{eq:4}), (\ref{eq:6}), together with these three facts.
 
Fig. \ref{fig:4}(a), (b), (c) and (d) show that the higher the bulk salt concentration, the narrower the voltage region where the permittivity is symmetric with respect to the surface voltage.  This is explained by the fact that at the higher bulk salt concentration, the number density of counterions at the neutral surface($\sigma=0$) is larger than that for the smaller bulk salt concentration.
Fig. \ref{fig:4}(a) also indicates that at high voltages the effect due to difference in bulk salt concentrations is negligible and the permittivity doesn't depend on the bulk salt concentration. This is attributed to the fact that at high voltages, the counterionic densities at the charged surface are equal to each other, regardless of the bulk salt concentration of the electrolyte solution as shown in Fig. \ref{fig:4}(c).

\begin{figure}
\includegraphics[width=1\textwidth]{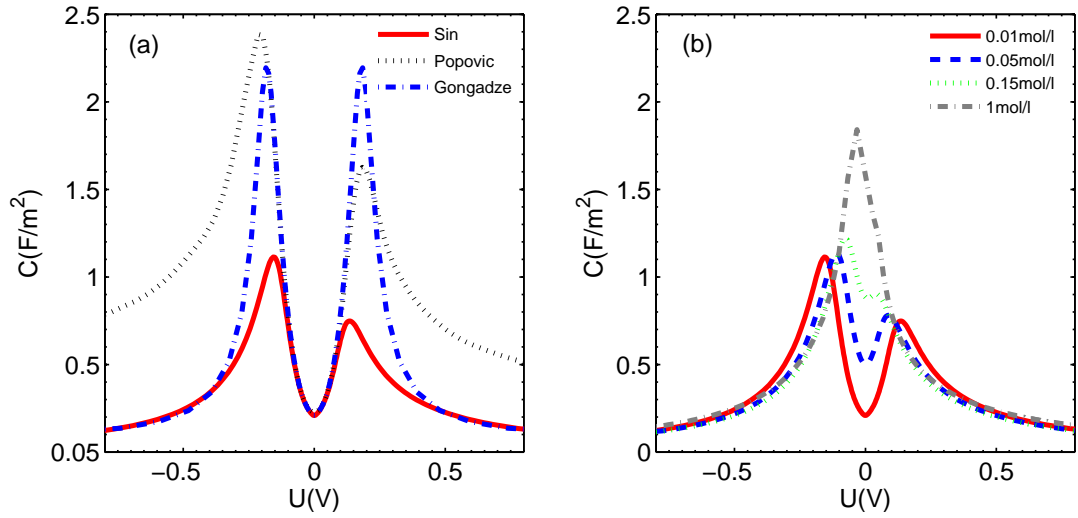}
\caption{(Color online) (a) The differential capacitance according to surface voltage for our approach( $V_{-}=2.21V_{+},V_{+}=0.15nm^3$), Popovic's approach( $V_{-}=2.21V_{+},V_{+}=0.15nm^3$) and Gongadze's approach($V_{-}=V_{+}=V_{w}$). Here $c_{0}/N_{A}=0.01mol/l$. (b) The differential capacitance according to surface voltage for our approach. The concentrations of the electrolyte solution are 0.01mol/l,0.05mol/l, 0.15mol/l, 1mol/l, respectively. Here $V_{-}=2.21V_{+},V_{+}=0.15nm^3$. Other parameters are the same as in Fig. \ref{fig:1}(a).}
\label{fig:5}
\end{figure}
Fig. \ref{fig:5}(a) shows the voltage dependences of differential capacitance calculated by using our approach , Popovic's approach\cite%
{Popovic_pre_2013} and Gongadze's approach\cite%
{Gongadze_bioelechem_2012}.  Fig. \ref{fig:5}(b) represents the voltage dependences of differential capacitance obtained by our approach for the following bulk salt concentrations; $c_{0}/N_{A}= 0.01mol/l, 0.05mol/l, 0.15mol/l, 1mol/l$.  In Fig. \ref{fig:5}(a) and Fig. \ref{fig:5}(b), other parameters are the same as in Fig. \ref{fig:1}(a). As shown in Fig. \ref{fig:5}(a) and Fig. \ref{fig:5}(b) all the differential capacitances aren't monotonic functions of surface voltage but show the same behavior, i.e. they first increase at low voltages, then have maxima at intermediate voltages and slowly decrease toward zero at higher voltages. 

In the low-voltage region, counterions can come closer to the charged surface easily by a removal of water molecules and the permittivity of electrolyte solution hardly changes.  In the high-voltage region, the packing and excluded volume effects become more important and the charge storage is more difficult, leading to a decrease of capacitance. Moreover, our approach predicts that water polarization lowers the permittivity of electrolyte solution at intermediate and high voltages and therefore enhances the decrease of differential capacitance.

Our approach provides the results which significantly differ from those in \cite%
{Popovic_pre_2013}. In particular, we note that the magnitude of the maximum differential capacitance for our approach is about half of that in \cite%
{Popovic_pre_2013}.
In the case of Popovic's approach \cite%
{Popovic_pre_2013}, the change of permittivity of the electrolyte solution due to the orientational ordering and excluded volume effect of water molecules is not taken into account which results in the overestimated magnitude of the maximum differential capacitance. In the case of our approach, the voltage for the maximum differential capacitance is slightly smaller than that for Popovic's approach. This is attributed to the fact that the excluded volume effect of water molecules prevents accumulation of ions close to the charged surface and causes the early onset of saturation of counterionic density. 

Gongadze's approach \cite%
{Gongadze_bioelechem_2012} accounts for the above-mentioned change of permittivity of electrolyte solution. However, Gongadze's approach is based on the assumption that all kinds of ions have the same size as water molecules have. As shown in Fig. \ref{fig:5}(a), the maximum differential capacitance for Gongadze's approach is larger than that for our approach. The reason for the discrepancy in the maximum capacitance between the two cases is that the larger size of counterions causes the early onset of saturation of counterionic density. 

At high voltages the differential capacitance for our approach is identical with one for Gongadze's approach as shown in Fig. \ref{fig:5}(a).  
The differential capacitance is defined as follows,
\begin{eqnarray}
	c=d\sigma/dU,
\label{eq:25}
\end{eqnarray}
 where $U=\psi\left(x=0\right)$ and $\sigma=\epsilon_{0}\epsilon_{r}\left(x=0\right)E\left(x=0\right)$. 
The larger size of counterion results in both the smaller permittivity of electrolyte solution and the larger magnitude of electric field strength as shown in Fig. \ref{fig:4}(a) and Fig. \ref{fig:4}(d).  These two effects cancel out each other at high voltages, so that  the variation of surface charge density with the surface voltage doesn't depend on volumes of ions. 

In the high voltage region, the differential capacitance for our approach varies with the surface voltage, regardless of the bulk salt concentration, as shown in Fig. \ref{fig:5}(b). This fact is explained by using Eq. (\ref{eq:25}), together with the fact that in the high voltage region,  the permittivity and the magnitude of the electric field strength doesn't depend on the bulk salt concentration as shown in Fig. \ref{fig:4}(a) and Fig. \ref{fig:4}(d).
 
Finally, we note that the differential capacitances for these three approaches are quite similar at low voltages because the electrolyte permittivity at the low voltages doesn't exhibit significant change as shown in Fig. \ref{fig:4}(a) and the excluded volume effect of ions is negligible. 

Although our differential capacitance curves represent some characteristics of experimental results of \cite%
{Grahame_amchem_1954, damaskin_jsolidstate_2011, foresti_jelecanal_1997},
 the comparison with experimental data is rather misleading. According to the Stern model \cite%
{Stern_zelektrochem_1924}, the total capacitance is given by an inner layer capacitance and a diffuse layer capacitance in series. For all practical cases the inner layer capacitance, which is due to the effect of electrode material and its coupling with the solution, is comparable to or smaller than the diffuse layer capacitance. Therefore, the diffuse layer capacitance is significantly higher than the total capacitance in practice. Because our theory predicts only the diffuse layer capacitance, our calculated capacitance is still higher than the experimental values. In order to  accurately determine properties of electric double layer, quantum mechanics should be used and it requires a lot of computational efforts as in \cite%
{badiali_jelecanal_1989, henderson_jchemphys_1985, schmickler_chemrev_1996, bonnet_prl_2013}. 
However, if we are interested in the properties of electric double layer related to only the electrolyte solution, our approach can predict the properties with a modest accuracy. Actually, our differential capacitance curves show behaviours similar to the results of density functional theory such as \cite%
{Jiang_chemphyslett_2011, frydel_jchemphys_2012}.

\section{Conclusions}
\large
We have presented the generalized Poisson-Boltzmann approach based on a lattice statistics taking into account the excluded volume effects of both ions and water molecules and the orientational ordering of water dipoles. 
Our approach takes into account also the influence of the sign of  surface charge density on electrostatic properties near the charged surface. 

It is predicted that the position of the minimum number density of water molecules can be more distant from the charged surface by an increase in volume of counterions and an increase in the surface charge density. 

The width of the voltage region in which electrostatic properties of electrolyte solution symmetrically vary with surface voltage, decreases with increasing bulk salt concentration.

 Additionally, the asymmetric, double-humped shape of differential capacitance curve compared to ones of \cite%
{Popovic_pre_2013, Gongadze_bioelechem_2012} shows early onset and lowering of the maximum of capacitance according to surface voltage which are attributed to not only the excluded volume effect of ions and water molecules but also the saturated orientational ordering of water dipoles.

\end{document}